\documentclass[preprint]{aastex}
\lefthead{Chambers and Murison}
\righthead{Pseudo-High-Order Symplectic Integrators}

\begin{document}
\title{Pseudo-High-Order Symplectic Integrators}

\author{J. E. Chambers}
\affil{Armagh Observatory, Armagh, Northern Ireland, BT61 9DG, United
Kingdom\\jec@star.arm.ac.uk}

\author{M. A. Murison}
\affil{Astronomical Applications Dept., U.S. Naval Observatory,
3450 Mass. Ave., NW, Washington D.C. 20392-5420\\murison@aa.usno.navy.mil}

\newcommand{\au}{{\sc au}}
\newcommand{\etal}{{\it et al.}}
\newcommand{\cf}{{\it c.f.}}
\newcommand{\eg}{{\it e.g.}}
\newcommand{\ie}{{\it i.e.}}

\begin{abstract}
Symplectic N-body integrators are widely used to study problems in
celestial mechanics. The most popular algorithms are of 2nd and 4th
order, requiring 2 and 6 substeps per timestep, respectively.
The number of substeps increases rapidly with order in timestep,
rendering higher-order methods impractical. However, symplectic
integrators are often applied to systems in which perturbations
between bodies are a small factor $\epsilon$ of the force due to a
dominant central mass. In this case, it is possible to create
optimized symplectic algorithms that require fewer substeps per
timestep. This is achieved by only considering error terms of order
$\epsilon$, and neglecting those of order $\epsilon^2$, $\epsilon^3$
etc. Here we devise symplectic algorithms with 4 and 6 substeps per
step which effectively behave as 4th and 6th-order integrators when
$\epsilon$ is small. These algorithms are more efficient than the
usual 2nd and 4th-order methods when applied to planetary systems.
\end{abstract}


\section{Introduction}
Symplectic integrators are widely used to study problems in celestial
mechanics. These integrators have two advantages over most other
algorithms. First, they exhibit no long-term build up in energy
error. Second, the motion of each object about the central body can
be ``built in'', so that the choice of step size, $\tau$, is
determined by the perturbations between bodies, whose magnitude is a
factor $\epsilon$ smaller than the forces due to the central body
(\cite{wis91}).

The most popular algorithm is the second-order symplectic integrator.
The error at each step is proportional to $\epsilon
\tau^3$, so that the likely error for an integration as a whole is
$\sim\epsilon \tau^2$. The second-order method is easy to implement,
consisting of only two substeps, including one force evaluation,
per time step. It is also very fast for integrations requiring
moderate accuracy.

For more accurate integrations, it is better to use the fourth-order
method (\cite{for90}). Here, the error at each step is proportional to
$\epsilon \tau^5$, although each step is computationally more expensive
since it consists of 6 substeps. Yoshida (1990) has developed 6th and 
8th-order symplectic integrators. However, these do not appear to
be competitive in most situations, due to the large number of
substeps required.

Here we show how to construct what are effectively high-order (4th, 6th
etc.) symplectic integrators that require fewer substeps per time
step than those in current use. The trick is to take into account the
dependence of each error term on $\epsilon$ when choosing the
coefficients for each substep. The algorithms are designed by
eliminating error terms proportional to $\epsilon$ up to the desired
order of the timestep. Error terms proportional to $\epsilon^2,
\epsilon^3$ etc., in low orders of the timestep, still exist. However,
in many situations these terms are negligible, and the integrators behave
as if they are of higher order than the leading error term in $\tau$
suggests.

Section~2 gives a quick review of how symplectic integrators are
traditionally constructed using Lie algebra. In Section~3, we show how
to build more efficient symplectic algorithms using fewer
substeps. Section~4 contains results of test integrations that compare
the new algorithms with traditional symplectic integrators.

\section{Symplectic Integrators}
Symplectic integrators for the N-body problem can be constructed
starting from Hamilton's equations of motion:
\begin{eqnarray}
\frac{dx_i}{dt}&=&\frac{\partial H}{\partial p_i} \nonumber \\
\frac{dp_i}{dt}&=&-\frac{\partial H}{\partial x_i}
\end{eqnarray}
where $x_i$ and $p_i$ are the coordinates and momenta of each body
respectively, and $H$ is the Hamiltonian of the system.

Using these equations, the rate of change of any dynamical quantity
$q({\bf x},{\bf p},t)$ can be expressed as
\begin{equation}
\frac{dq}{dt}=\sum_{i=1}^{3N} \left(\frac{\partial q}{\partial x_i}
\frac{\partial H}{\partial p_i}-\frac{\partial q}{\partial p_i}
\frac{\partial H}{\partial x_i}\right)\equiv\{q,H\}\equiv Fq
\label{eq20}
\end{equation}
where $\{,\}$ are Poisson brackets, and $F$ is a differential operator.

The formal solution of equation (\ref{eq20}) is
\[
q(t)=e^{\tau F}q(t-\tau)
    =\left(1+\tau F+\frac{\tau ^2F^2}{2}+\ldots\right)q(t-\tau)
\]
where $F^2q = F(Fq)$ etc.

Now suppose that we are able to split the Hamiltonian into two pieces,
$H_A$ and $H_B$, so that each part of the problem can be solved relatively
easily in the absence of the other. The solution for $q$ becomes
\begin{equation}
q(t)=e^{\tau(A+B)} q(t-\tau)
\label{eq30}
\end{equation}
where $A$ and $B$ are differential operators related to $H_A$ and
$H_B$ respectively, in the same way that $F$ is related to $H$.

The Baker-Campbell-Hausdorff (BCH) formula states that, for any
noncommutative operators $A$ and $B$,
\[
\exp(A)\cdot\exp(B)=\exp(C)
\]
where $C$ is a series consisting of nested commutators,
\[
C=A+B+\frac{1}{2}[A,B]+\frac{1}{12}[A,A,B]+\frac{1}{12}[B,B,A]+\cdots
\]
where the commutator $[A,B]=AB-BA\neq 0$ in general
(see, for example, Yoshida 1990 or Forest and Ruth 1990). Here, we
have used the nested commutator notation $[A,B,C]=[A,[B,C]]$, etc..

Hence, if we evolve $q$ under the two parts of the Hamiltonian
separately, one after the other, we have
\begin{equation}
\exp(\tau A)\cdot\exp(\tau B)q(t-\tau)=
\exp\left[\tau F+\frac{\tau^2}{2}[A,B]+\cdots\right]q(t-\tau)
\label{eq40}
\end{equation}
This is identical to the righthand side of
equation (\ref{eq30}) to $O(\tau)$, and so equation
(\ref{eq40}) represents a first-order integrator. Each step of the
integrator consists of 2 substeps, with the whole step giving an error of
$O(\tau^2)$. Alternatively, we can say that the integrator exactly
solves a problem whose Hamiltonian is given by
\[
H_{integ}=H+\frac{\tau}{2}\{H_B,H_A\}+O(\tau^2)
\]
(see, for example, \cite{sah92}). Provided that $\tau$ is
small, and $\{H_B,H_A\}$ remains bounded, the energy of the
integrated system will always be near to that of the real system.

Other integrators can be found by combining exponential
operators in such a way that they are equivalent to equation
(\ref{eq30}) up to a given order in $\tau$. For example, we have the
second-order symplectic integrator
\begin{eqnarray*}
{\rm S2A}&=&
\exp\left(\frac{\tau}{2}A\right)\cdot\exp(\tau B)\cdot
\exp\left(\frac{\tau}{2}A\right) \\
&=&\exp\left[\tau F+\frac{\tau^3}{12}[B,B,A]-\frac{\tau^3}{24}[A,A,B]
+\ldots\right]
\end{eqnarray*}
When many integration steps are performed one after another,
the $\exp(\tau A/2)$ terms at the end of one step and the start of
another can be combined. Hence, the second-order integrator
also consists of only 2 substeps, except at the beginning and the end
of an integration.

Another second-order integrator is
\begin{eqnarray}
{\rm S2B}&=&
\exp\left(\frac{\tau}{2}B\right)\cdot\exp(\tau A)\cdot
\exp\left(\frac{\tau}{2}B\right) \nonumber \\
&=&\exp\left[\tau F+\frac{\tau^3}{12}[A,A,B]-\frac{\tau^3}{24}[B,B,A]
+\ldots\right]
\label{eq70}
\end{eqnarray}
The distinction between S2A and S2B (which at first sight appear to be
the same) will become apparent in the next section, when we consider
situations in which $A$ and $B$ are of different magnitude.

Forest and Ruth (1990) give a fourth-order symplectic integrator with
6 substeps per step:
\begin{eqnarray*}
{\rm S4B}&=&
\exp\left(\frac{\tau B}{2c}\right)\cdot
\exp\left(\frac{\tau A}{c}\right)\cdot
\exp\left[\frac{\tau B(1-k)}{2c}\right]\cdot
\exp\left(\frac{-\tau kA}{c}\right)\cdot
\exp\left[\frac{\tau B(1-k)}{2c}\right] \\
&\cdot&
\exp\left(\frac{\tau A}{c}\right)\cdot
\exp\left(\frac{\tau B}{2c}\right) \\
&=&\exp[\tau F+O(\tau^5)]
\end{eqnarray*}
where $k=2^{1/3}$ and $c=2-k$. Note that the middle 3 substeps move in the
opposite direction to the integration as a whole.

Higher-order integrators require progressively more substeps. Yoshida
(1990) gives examples of 6th and 8th-order integrators using 14 and 30
substeps respectively. In the next section, we will show how to
create what are {\it effectively\/} 4th and 6th order integrators (and in
principle, 8th-order etc.) using fewer substeps than are required
conventionally.

\section{Constructing Pseudo-Order Integrators}
Up to this point we have not considered the details of how $H$ is
split. Suppose that one part of the Hamiltonian is much smaller
than the other, \ie\ $H=H_A+\epsilon H_B$, where $\epsilon\ll 1$.
Now consider the error terms in the second-order integrator of
equation (\ref{eq70}):
\[
{\rm S2B}=\exp\left[\tau F+\frac{\epsilon\tau^3}{12}[A,A,B]
-\frac{\epsilon^2\tau^3}{24}[B,B,A]+\cdots\right]
\]
One of the $O(\tau^3)$ error terms is smaller than the other
by a factor of $\epsilon$. 

Similarly, for the fourth order integrator:
\[
{\rm S4B}=\exp[\tau F+O(\epsilon\tau^5)+O(\epsilon^2\tau^5)
+O(\epsilon^3\tau^5)+O(\epsilon^4\tau^5)]
\]
Some of these error terms are insignificant compared to others, and
yet this was not taken into account when constructing the
integrator. The only design criterion was that S4B should contain no
error terms below the fifth power in the timestep. If we take
into account the dependence of the error terms on both $\tau$ and
$\epsilon$, we can design more efficient symplectic integrators.

%The derivations below involve some rather tedious algebra, however the
%results are simple. If you are not interested in the derivations,
%you can go directly to Section~3.1 for the results.

To construct the new integrators, we again employ the BCH
formula. Adapting the expression for the BCH
formula given by Yoshida (1990), we have:
\begin{eqnarray*}
&&\exp(a_1\tau A)\cdot\exp(b_1\epsilon\tau B) \\
&=&\exp\left[(a_1A+\epsilon b_1B)\tau
+\epsilon\tau^2\left(\frac{a_1b_1}{2}\right)       [A,B]
+\epsilon\tau^3\left(\frac{a_1^2b_1}{12}\right)    [A,A,B]
+\epsilon^2\tau^3\left(\frac{a_1b_1^2}{12}\right)  [B,B,A]\right. \\
&+&\left.\epsilon^2\tau^4\left(\frac{a_1^2b_1^2}{24}\right) [A,B,B,A]
-\epsilon\tau^5\left(\frac{a_1^4b_1}{720}\right)   [A,A,A,A,B]
-\epsilon^4\tau^5\left(\frac{a_1b_1^4}{720}\right) [B,B,B,B,A]
+\cdots\right]
\end{eqnarray*}
where $a_1$ and $b_1$ are constants. Additional fifth-order
commutators are present; however, we will only require terms that
contain either $A$ or $B$ once, since these are the type of
error term we are seeking to eliminate.

Applying the BCH formula twice, Yoshida (1990) gives an expression for
a symmetric product of three exponential operators:
\begin{eqnarray}
&&\exp(b_1\epsilon\tau B)\cdot\exp(a_1\tau A)\cdot
  \exp(b_1\epsilon\tau B)
\nonumber \\
&=&\exp\left[(a_1A+2\epsilon b_1B)\tau
+\epsilon\tau^3\left(\frac{a_1^2b_1}{6}\right)      [A,A,B]
-\epsilon^2\tau^3\left(\frac{a_1b_1^2}{6}\right)    [B,B,A]\right.
\nonumber \\
&-&\left.\epsilon\tau^5\left(\frac{a_1^4b_1}{360}\right) [A,A,A,A,B]
+\epsilon^4\tau^5\left(\frac{7a_1b_1^4}{360}\right) [B,B,B,B,A]
+\cdots\right]
\label{eq100}
\end{eqnarray}
Again we have neglected fifth-order terms that contain both $A$ and
$B$ more than once. Note that there are no terms containing even
powers of the timestep: Yoshida shows that this is a general property
of any symmetric arrangement of exponential operators. From now on we
will consider only symmetrical integrators because of this property.

We need to extend equation (\ref{eq100}) once more to get a
pseudo-fourth order integrator, and twice more for a pseudo-sixth
order one. By substituting $a_2A$ for $b_1B$ in equation
(\ref{eq100}), and substituting the righthand side of equation
(\ref{eq100}) for $a_1A$, we get:
\begin{eqnarray}
&&\exp(a_2\tau A)\cdot\exp(b_1\epsilon\tau B)\cdot\exp(a_1\tau A)
\cdot\exp(b_1\epsilon\tau B)\cdot\exp(a_2\tau A) 
\nonumber\\ 
&=&\exp\left\{(a_1+2a_2)\tau A+2b_1\epsilon\tau B
+\epsilon\tau^3\left(\frac{b_1}{6}\right)[(a_1+2a_2)^2-6a_2(a_1+a_2)]
\,[A,A,B]\right.
\nonumber \\
&+&\epsilon^2\tau^3\left(\frac{b_1^2}{6}\right)(4a_2-a_1)\,[B,B,A]
-\epsilon\tau^5\left(\frac{b_1}{360}\right)
[(a_1+2a_2)^4-30a_2^2(a_1+a_2)^2]\,[A,A,A,A,B]
\nonumber \\
&-&\left.\epsilon^4\tau^5\left(\frac{b_1^4}{360}\right)
(16a_2-7a_1)\,[B,B,B,B,A]+\cdots\right\}
\label{eq110}
\end{eqnarray}

Finally, substituting the righthand side of equation (\ref{eq110}) for
$a_1A$ in equation (\ref{eq100}), and replacing $b_1B$ with $b_2B$, we
arrive at
\begin{eqnarray}
&&\exp(b_2\epsilon\tau B)\cdot\exp(a_2\tau A)\cdot
  \exp(b_1\epsilon\tau B)\cdot\exp(a_1\tau A)\cdot
  \exp(b_1\epsilon\tau B)\cdot\exp(a_2\tau A)\cdot
  \exp(b_2\epsilon\tau B) \nonumber\\ 
&=&\exp\left\{(a_1+2a_2)\tau A+2(b_1+b_2)\epsilon\tau B
+\epsilon\tau^3\left[\frac{(b_1+b_2)(a_1+2a_2)^2-6a_2b_1(a_1+a_2)}{6}
\right][A,A,B]\right.
\nonumber \\
&-&\epsilon^2\tau^3\left[\frac{(a_1+2a_2)(b_1+b_2)^2-6a_2b_1^2}{6}\right]
[B,B,A]
\nonumber \\
&-&\epsilon\tau^5\left[\frac{(b_1+b_2)(a_1+2a_2)^4-30a_2^2b_1(a_1+a_2)^2}
{360}\right] [A,A,A,A,B]
\nonumber \\
&+&\left.\epsilon^4\tau^5\left[\frac{7(a_1+2a_2)(b_1+b_2)^4
-60a_2b_1^2(b_1+b_2)^2+30a_2b_1^4}{360}\right] [B,B,B,B,A]+\cdots\right\}
\label{eq120}
\end{eqnarray}

The first stage in converting these general expressions into
specific integrators is to make the coefficients of the linear $A$
and $B$ terms equal to 1. This places two constraints on the values of
the constants. We can then get what is effectively a
4th-order integrator by simply eliminating the $[A,A,B]$ term from
equation (\ref{eq110}). The leading error terms will now be
$O(\epsilon^2\tau^3)$ and $O(\epsilon\tau^5)$.
Provided that $\epsilon$ is small enough, the largest error term will
be $O(\epsilon\tau^5)$, and the integrator effectively will be of
fourth order in the timestep. Applying these conditions, we require
\begin{eqnarray}
a_1+2a_2&=&1 \nonumber \\
2b_1&=&1 \nonumber \\
1-6a_2(1-a_2)&=&0
\label{eq150}
\end{eqnarray}
where we have used the first two of equations (\ref{eq150}) in
deriving the third.

Alternatively, we may construct an integrator in which each step begins by
advancing $H_B$ instead of $H_A$. Unlike conventional symplectic
integrators, such as $S2A$ and $S2B$, we cannot use the same set of
coefficients when exchanging $A$ and $B$. Instead, we must derive a
new set of coefficients by interchanging $A$ and $\epsilon B$ in
equation (\ref{eq110}) and then eliminating the new $[A,A,B]$
term. When we do this, the first two of equations (\ref{eq150}) remain as
before, but the third expression becomes
\begin{equation}
6a_2-1=0
\label{eq160}
\end{equation}

To get a pseudo-6th-order integrator, we eliminate terms 
containing $[A,A,B]$ and $[A,A,A,A,B]$. This will produce
an extra constraining equation, so we need an extra constant. We
get this by using an integrator with the form of equation (\ref{eq120}) 
instead of equation (\ref{eq110}). The corresponding equations for the
constants are 
\begin{eqnarray}
a_1+2a_2&=&1 \nonumber \\
2(b_1+b_2)&=&1 \nonumber \\
1/2-6a_2b_1(1-a_2)&=&0 \nonumber \\
1/2-30a_2^2b_1(1-a_2)^2&=&0
\label{eq170}
\end{eqnarray}

If we prefer an integration step that begins by advancing $H_A$, we
can interchange $A$ and $\epsilon B$ in equation (\ref{eq120}), and
eliminate the new $[A,A,B]$ and $[A,A,A,A,B]$ terms. In this case, the
last two of equations (\ref{eq170}) become
\begin{eqnarray}
1/4-6a_2b_1^2&=&0 \nonumber \\
7/16-15a_2b_1^2+30a_2b_1^4&=&0
\label{eq180}
\end{eqnarray}

The leading error terms for each of these integrators are 
$O(\epsilon^2\tau^3)$ and $O(\epsilon\tau^7)$. The latter will be
dominant if $\epsilon$ is small enough, so that the algorithms behave
as 6th-order integrators.

\subsection{Pseudo-4th and 6th-Order Examples}
Solving equations (\ref{eq150}) and (\ref{eq160}), we obtain two
pseudo-4th-order integrators:
\begin{eqnarray*}
{\rm S4A*}&=&\exp\left[\frac{\tau A}{2}\left(1-\frac{1}{\surd 3}\right)\right]
\cdot\exp\left(\frac{\epsilon\tau B}{2}\right)
\cdot\exp\left(\frac{\tau A}{\surd 3}\right)
\cdot\exp\left(\frac{\epsilon\tau B}{2}\right)
\cdot\exp\left[\frac{\tau A}{2}\left(1-\frac{1}{\surd 3}\right)\right] \\
&=&\exp\left[\tau F
+\epsilon^2\tau^3\left(\frac{2-\surd 3}{24}\right)[B,B,A]
-\frac{\epsilon\tau^5}{4320}[A,A,A,A,B]+\cdots\right] \\
&& \\
{\rm S4B*}&=&\exp\left(\frac{\epsilon\tau B}{6}\right)
\cdot\exp\left(\frac{\tau A}{2}              \right)
\cdot\exp\left(\frac{2\epsilon\tau B}{3}     \right)
\cdot\exp\left(\frac{\tau A}{2}              \right)
\cdot\exp\left(\frac{\epsilon\tau B}{6}      \right) \\
&=&\exp\left[\tau F
+\frac{\epsilon^2\tau^3}{72}[B,B,A]
+\frac{\epsilon\tau^5}{2880}[A,A,A,A,B]+\cdots\right]
\end{eqnarray*}
where the asterisk in S4A* indicates that it only behaves as a 4th
order integrator for certain values of~$\tau$.

Equations~\ref{eq170} and \ref{eq180} give two pseudo-6th-order integrators:
\begin{eqnarray*}
{\rm S6A*}&=&\exp\left[\frac{\tau A}{2}\left(1-\frac{3}{\surd
15}\right)\right]
\cdot\exp\left(\frac{5\epsilon\tau B}{18}\right)
\cdot\exp\left(\frac{3\tau A}{2\surd 15}\right)
\cdot\exp\left(\frac{4\epsilon\tau B}{9}\right)
\cdot\exp\left(\frac{3\tau A}{2\surd 15}\right) \\
&\cdot&\exp\left(\frac{5\epsilon\tau B}{18}\right)
\cdot\exp\left[\frac{\tau A}{2}\left(1-\frac{3}{\surd 15}\right)\right] \\
&=&\exp\left[\tau F+\epsilon^2\tau^3\left(\frac{54-13\surd
15}{648}\right) [B,B,A] +O(\epsilon\tau^7)\right] \\
&& \\
{\rm S6B*}&=&\exp\left(\frac{\epsilon\tau B}{12}\right)
\cdot\exp\left[\frac{\tau A}{2}\left(1-\frac{1}{\surd 5}\right)\right]
\cdot\exp\left(\frac{5\epsilon\tau B}{12}\right)
\cdot\exp\left(\frac{\tau A}{\surd 5}\right)
\cdot\exp\left(\frac{5\epsilon\tau B}{12}\right) \\
&\cdot&\exp\left[\frac{\tau A}{2}\left(1-\frac{1}{\surd 5}\right)\right]
\cdot\exp\left(\frac{\epsilon\tau B}{12}\right) \\
&=&\exp\left[\tau F+\epsilon^2\tau^3\left(\frac{13-5\surd
5}{288}\right) [B,B,A] +O(\epsilon\tau^7)\right]
\end{eqnarray*}

Unlike the 4th-order algorithm of Forest and Ruth (1990) and the
6th-order integrators of Yoshida (1990), the algorithms above contain
no substeps that move in the opposite direction to the main integration.
An additional solution exists for each of equations (\ref{eq150}),
(\ref{eq170}) and (\ref{eq180}), however these have error terms with
larger numerical coefficients than the integrators we show here.

The same method can be used to generate a pseudo-8th-order integrator
and so on. Each higher order will require only 2 more substeps than
the previous one, since only one more commutator needs to be eliminated in
each case. For example, to create a pseudo-8th-order integrator
requires the elimination of the $[A,A,A,A,A,A,B]$ term in addition to
those that are absent from the pseudo-6th-order case.
However, depending on the system to be integrated, there
will come a point at which the $\epsilon^2\tau^3$ error term becomes
the most important. In principle, one could devise another set of
integrators that eliminates terms in $\epsilon^2\tau^m$ for small $m$,
in addition to terms in $\epsilon\tau^m$. However, achieving each new
order will generally require the elimination of more than one
commutator term, so that these integrators increase in complexity
much more rapidly than those described here.

Murison and Chambers (1999) have independently derived the two
4th-order integrators above, among others, using a symbolic algebra
package. Further results from that approach will follow in another paper.
We note that the pseudo-order algorithms can be adapted to
use independent timesteps for each planet (\cf\ \cite{sah94}),
or to include close encounters (\cite{dun98,cha99}).

\section{Numerical Comparisons}
In this section, we test the pseudo-4th and 6th order integrators
derived in Section~3 against the well-known 2nd and 4th-order
symplectic algorithms. We use the ``mixed-variable'' method of
Wisdom and Holman (1991), in which the Hamiltonian is divided into
a Keplerian part, $H_K$, and an interaction part, $H_I$. Under $H_K$,
each object moves on an unperturbed Keplerian orbit about the central
body. Under $H_I$, each object remains fixed while receiving an
impulse due to the gravitational perturbations of all the other
objects except the central body. As suggested by Wisdom and Holman, we
use Jacobi coordinates rather than barycentric coordinates. The
integrations themselves were carried out using a modified version of
the {\it Mercury\/} integrator package (\cite{cha97}).

\begin{figure}
\plottwo{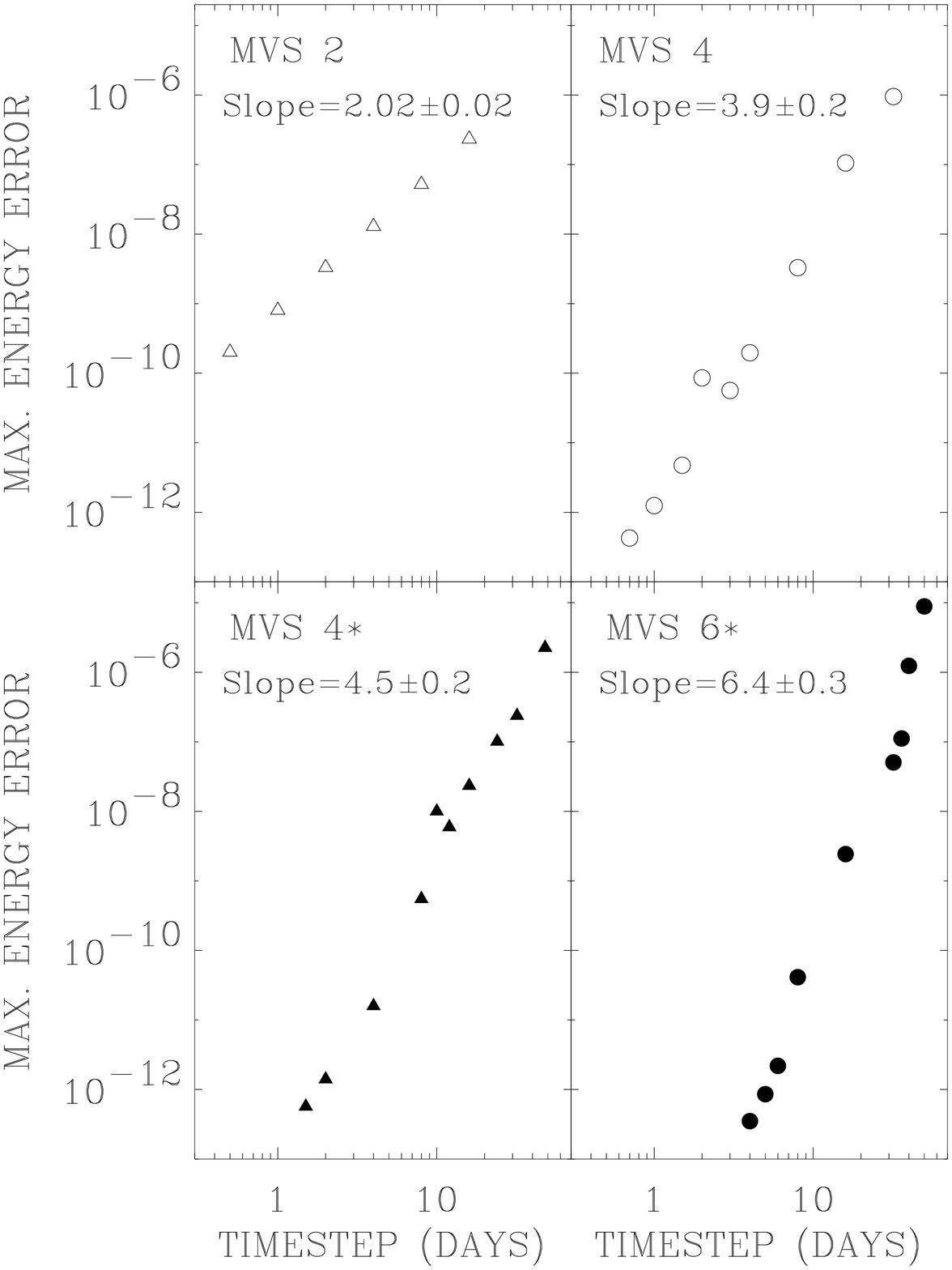}{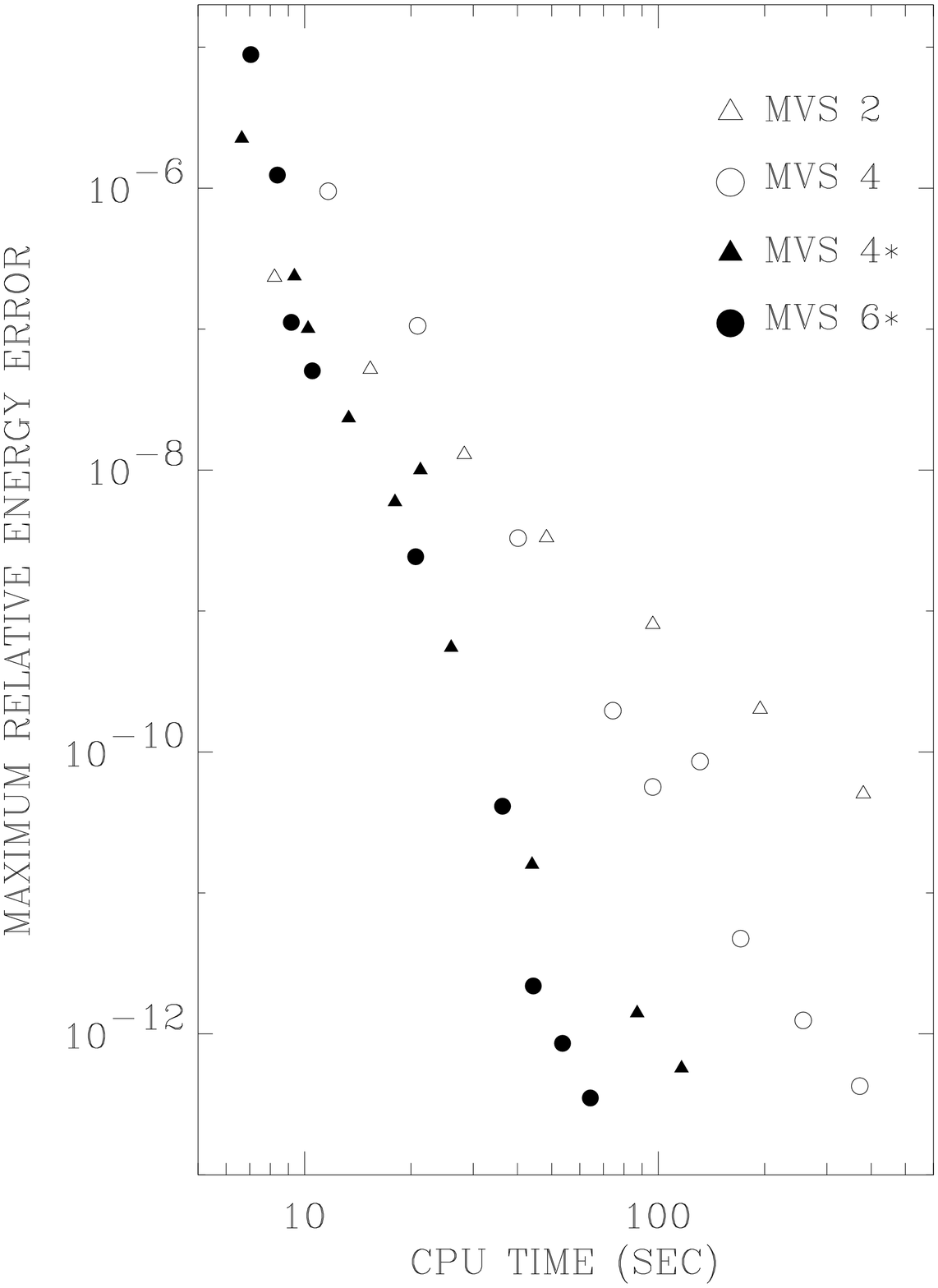}
\caption{Maximum relative energy error versus step size for a
10000-year integration of the 4 terrestrial planets using various
symplectic integrators.}
\caption{Maximum relative energy error 
versus CPU time for a 10000-year integration of the 4 terrestrial 
planets using various symplectic integrators.}
\end{figure}

\begin{figure}
\plottwo{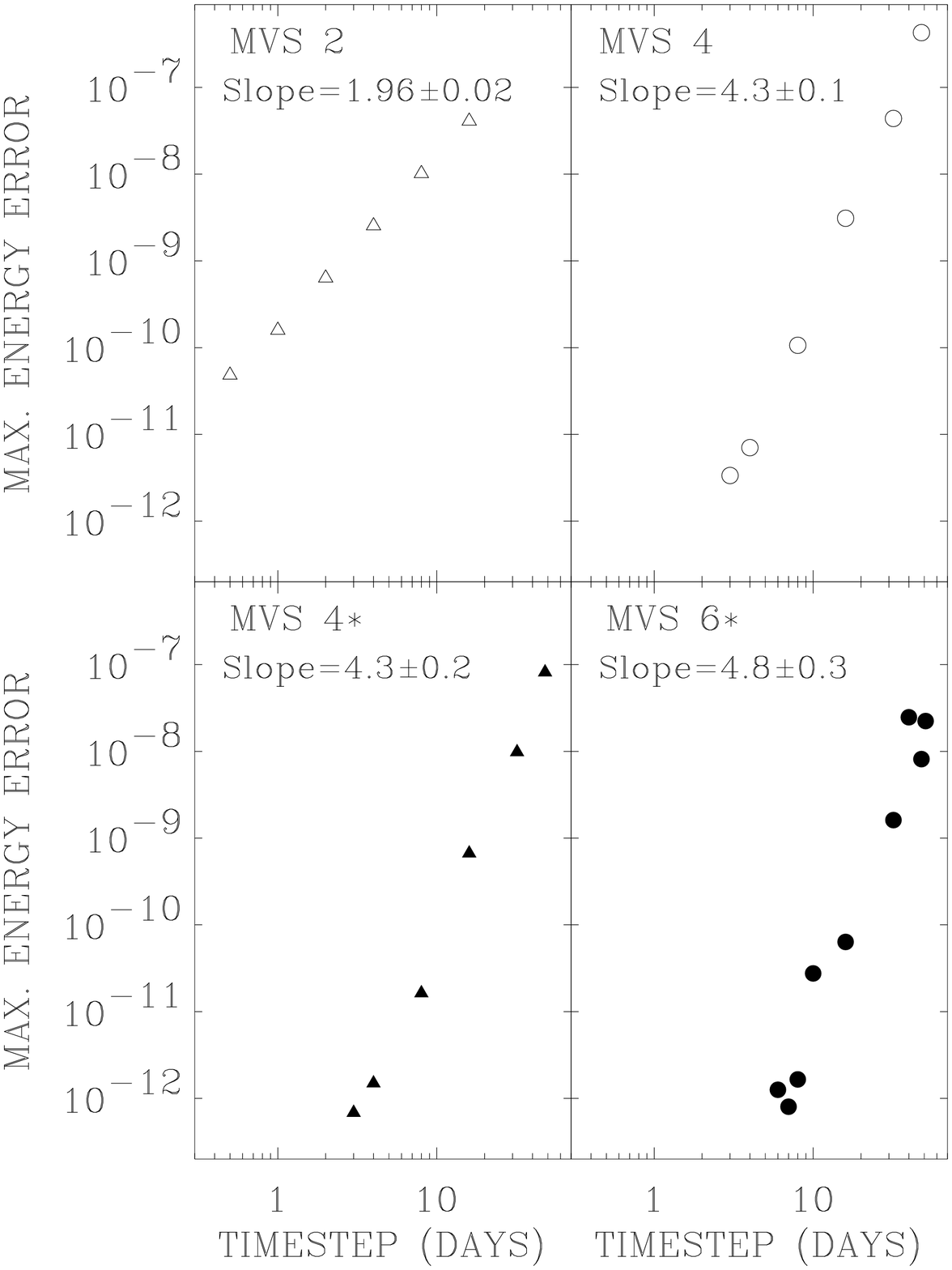}{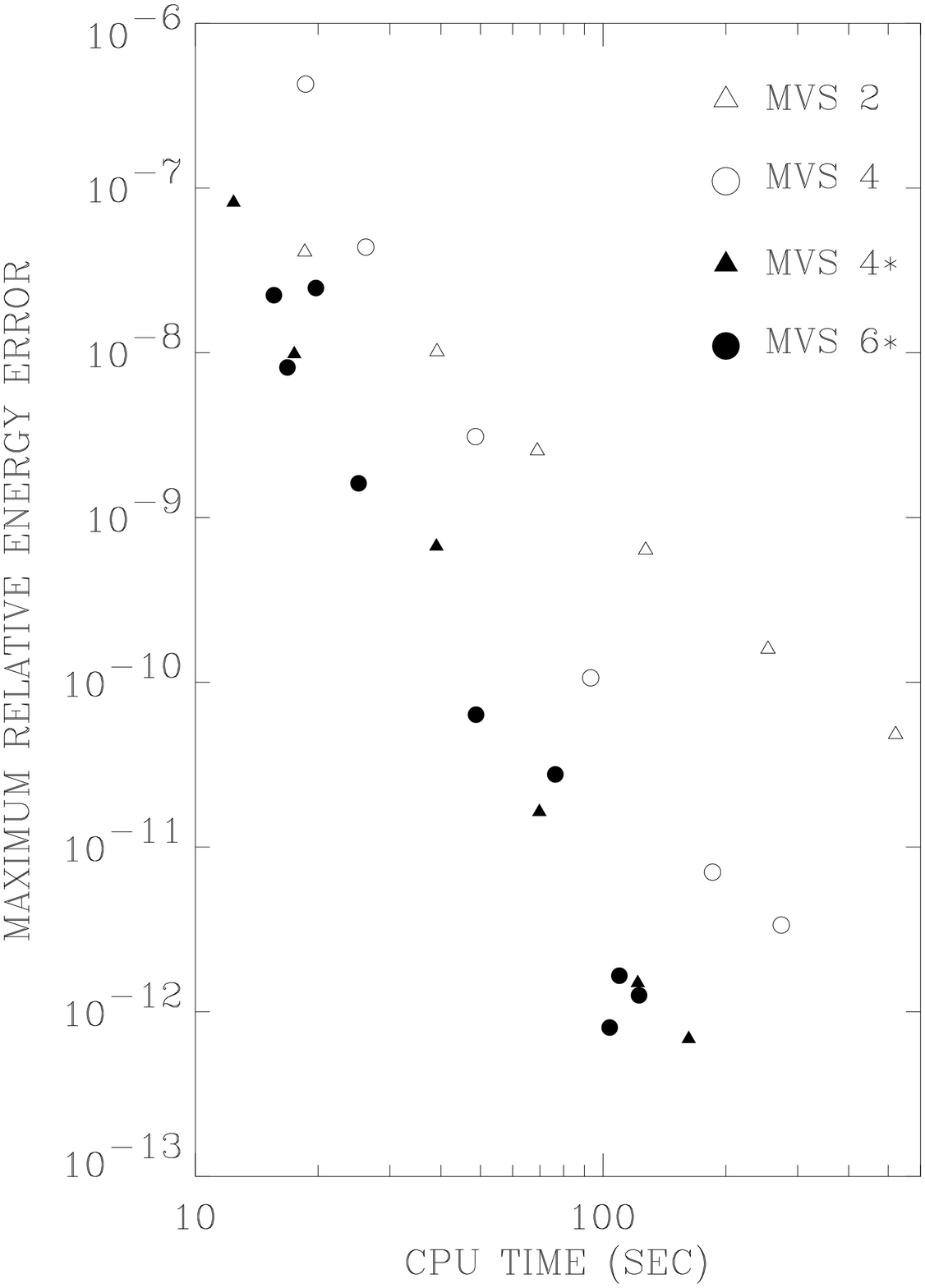}
\caption{Maximum relative energy error versus step size for a
10000-year integration of the 9 planets using various
symplectic integrators.}
\caption{Maximum relative energy error versus 
CPU time for a 10000-year integration of the 9 planets using various 
symplectic integrators.}
\end{figure}

The pseudo-order integrators require that the ratio
$\epsilon=H_I/H_K\ll 1$. In our first test, we integrate the
orbits of the 4 inner planets of the solar system in the absence of
the outer planets. In this case $\epsilon\sim 10^{-5}$. Figure~1 shows
the results of a 10000-year integration using the conventional 2nd and
4th-order symplectic integrators, S2B and S4B, and the pseudo-order
integrators S4B* and S6B*. For each integration, the maximum relative
energy error is shown as a function of the step size.

For the 2nd and 4th-order integrators, the maximum energy error is
roughly proportional to $\tau^2$ and $\tau^4$ respectively, where
$\tau$ is the timestep. This is what we would expect to find. For the
pseudo-4th and 6th-order integrators, the maximum energy error varies
as $\tau^4$ and $\tau^6$. That is, they behave as 4th and 6th-order
integrators, as we anticipated, despite the fact that they contain
error terms of lower order in the timestep.

Using the mean relative energy error per integration instead of
the maximum error gives results similar to Figure~1. The corresponding
slopes are $2.10\pm 0.05$ for S2B, $3.9\pm 0.3$ for S4B, $4.6\pm 0.3$
for S4B* and $6.4\pm 0.4$ for S6B*.

Figure~2 shows the amount of CPU time required for the integrations
shown in Figure~1. For energy errors of 1 part in $10^6$ or $10^8$
there is not much to choose between the four algorithms. For higher
levels of accuracy, S4B outperforms S2B. However, the pseudo
integrators S4B* and S6B* do even better. At an accuracy of 1 part in
$10^{10}$, they are roughly an order of magnitude faster than the
conventional second-order integrator, and 3 times faster than the 4th-order
integrator. For accuracies of better than $10^{-11}$, S6B*
shows greater performance than S4B*.

The pseudo-4th order integrator is more efficient than the real 4th-order
integrator for two reasons. It requires fewer substeps per
time step, and it has a slightly smaller leading error term.

%\begin{figure}
%\figurenum{4}
%\plotone{pseudo_fig4.eps}
%\epsscale{0.5}
%%\psfig{file=pseudo_fig4.epsi,height=210mm,clip=,angle=0}
%\caption{Maximum relative energy error versus CPU time for a 10000-year
%integration of the 9 planets using various symplectic integrators.}
%\end{figure}

As a more interesting test, we integrated the whole planetary system
(Mercury to Pluto) for 10000 years. Figure~3 shows the energy-error results
of these integrations. The behaviour of S2B, S4B and S4B* is similar
to that for the integrations of the terrestrial planets. However, the
energy error for S6B* varies roughly as $\tau^5$ rather than $\tau^6$.
It is not obvious why this should be, although the difference from the
terrestrial-planet integration (Figure~1) is presumably due to the fact that
$\epsilon$ is two orders of magnitude larger in this case.

Figure~4 shows the CPU time required for the integrations of the 9 planets.
The results are similar to the integration of the inner
planets, except that S6B* has only a marginal advantage over S4B* at
the highest levels of accuracy.

Since writing the original draft of this manuscript, we have become
aware of the symplectic corrector method of Wisdom \etal\ (1996),
which substantially improves the efficiency of the second-order
symplectic integrator. We present the pseudo-order integrators
as an alternative strategy for designing accurate algorithms.
It is possible to devise other symplectic correctors
using the same approach we use in Section~3 to design the integrator kernel:
that is, by considering the dependence of the resulting error
terms on $\epsilon$ as well as $\tau$ (\cite{mik97,rau99}).
Finally we suggest that it may be possible to design symplectic
correctors to improve the performance of pseudo-order algorithms,
since the pseudo-order methods exhibit similar high-frequency
oscillations in energy error to the second and 4th-order symplectic
integrators (see Figure~5).

%\vskip 4in
%\hspace{1in}
%\vskip 3.75in
%\special{ps: pseudo_fig5.eps x=3.5in y=3.5in}

\begin{figure}
\figurenum{5}
\plotone{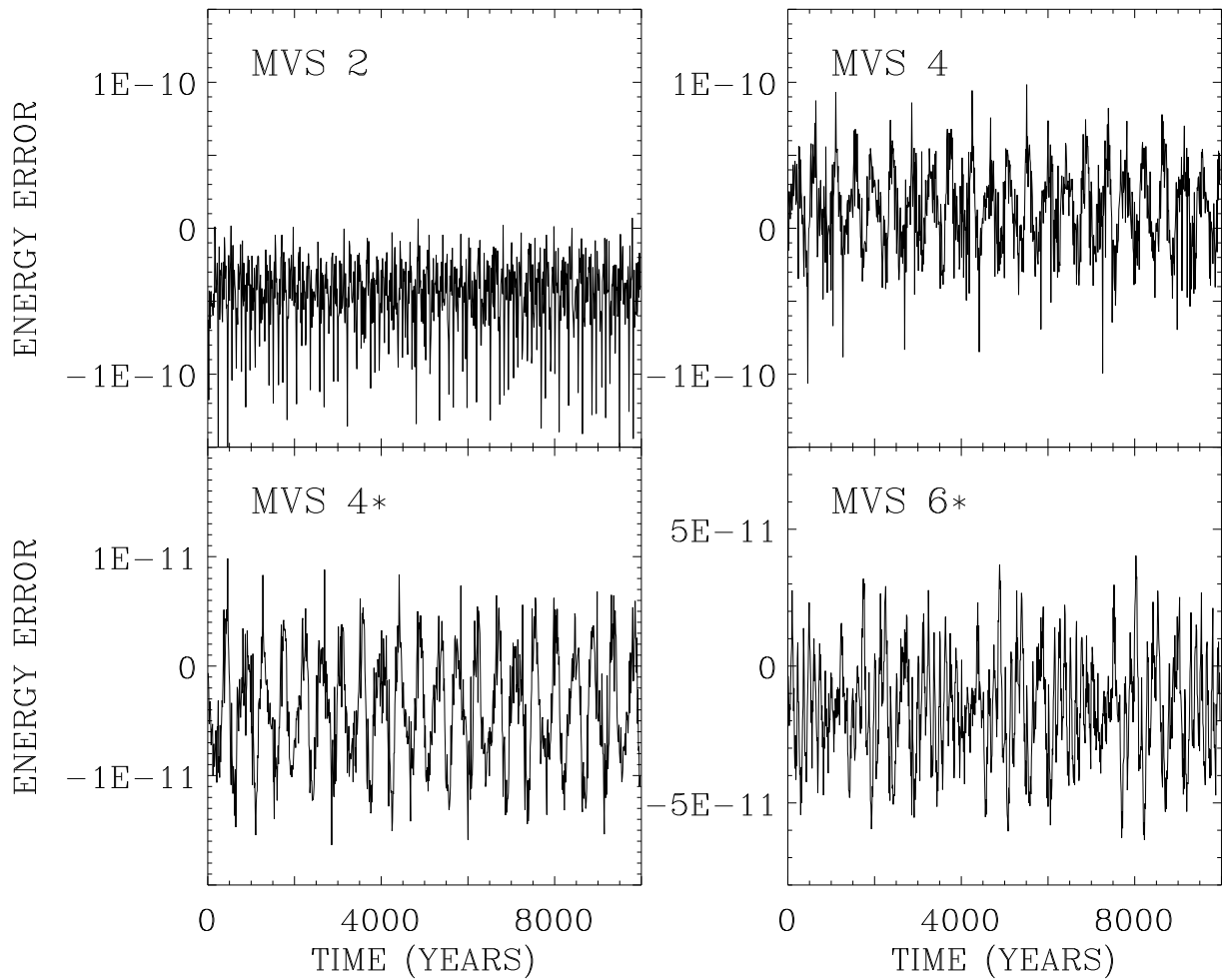}
\caption{Relative energy error versus time for 10000-year
integrations of the 9 planets using various symplectic integrators.}
\end{figure}

In summary, we conclude that the new pseudo-order integrators
outperform the widely-used 2nd and 4th-order algorithms at all
reasonable values of the energy error, for problems involving a
dominant central mass.

\acknowledgments
Research at Armagh Observatory is grant-aided by the Dept. of
Education, Northern Ireland. The test integrations described in this
paper were carried out using computers purchased on a PPARC research
grant.

\newpage
\section*{Figure Captions}
\begin{description}
\item[Figure 1:] Maximum relative energy error versus step size for a
10000-year integration of the 4 terrestrial planets using various
symplectic integrators.

\item [Figure 2:] Maximum relative energy error versus CPU time for a
10000-year integration of the 4 terrestrial planets using various
symplectic integrators.

\item [Figure 3:] Maximum relative energy error versus step size for a
10000-year integration of the 9 planets using various
symplectic integrators.

\item[Figure 4:] Maximum relative energy error versus CPU time for a
10000-year integration of the 9 planets using various symplectic
integrators.

\item[Figure 5:] Relative energy error versus time for 10000-year
integrations of the 9 planets using various symplectic integrators.

\end{description}

\end{document}